\documentclass[11pt]{article} 
\usepackage{amssymb} 
\usepackage{amsmath} 
 
\usepackage{graphicx} 
 
\def\beq{\begin{eqnarray}}    %%%  begequation/eqnarray 
\def\eeq{\end{eqnarray}}      %%%  endequation/eqnarray 
 
\def\Box{\square}                       %%% Box 
\def\tr{\,\mbox{tr}\,}                  %%% trace 
\def\al{\alpha} 
\def\be{\beta} 
 
\def\ga{\gamma} 
\def\de{\delta}

\def\ze{\zeta}

\def\la{\lambda} 
\def\na{\nabla} 
\def\pa{\partial} 
 
\def\si{\sigma} 
 
\def\ph{\varphi}

\def\Ga{\Gamma} 
\def\De{\Delta} 
\def\La{\Lambda}

\catcode`\@=11 
 
\newif\if@fewtab\@fewtabtrue 
 
\def\eqnarray{\def\draftnote{{}}\global\@fewtabtrue 
\stepcounter{equation}\let\@currentlabel=\theequation 
\global\@eqnswtrue 
\global\@eqcnt\z@\tabskip\@centering\let\\=\@eqncr 
$$\halign to \displaywidth\bgroup\@eqnsel\hskip\@centering\@eqcnt\z@ 
  $\displaystyle\tabskip\z@{##}$&\global\@eqcnt\@ne 
  \hskip 1\arraycolsep \hfil$\displaystyle{##}$\hfil 
  &\global\@eqcnt\tw@ \hskip 1\arraycolsep 
$\displaystyle\tabskip\z@{##}$ 
\hfil  \tabskip\@centering&\global\@eqcnt\thr@@\llap{##}\tabskip\z@ 
\cr} 
 
\def\endeqnarray{\@@eqncr\egroup 
      \global\advance\c@equation\m@ne$$\global\@ignoretrue} 
 
{\count255=\time\divide\count255 by 60 
\xdef\hourmin{\number\count255} 
\multiply\count255 by-60\advance\count255 by\time 
\xdef\hourmin{\hourmin:\ifnum\count255<10 0\fi\the\count255}} 
\def\tilde{\widetilde} 
\def\hat{\widehat} 
\def\pref#1{(\ref{#1})} 
 
\newif\if@fewtab\@fewtabtrue 
\def\ceqno{\global\@fewtabfalse 
    \ifcase\@eqcnt \def\@tempa{& & &}\or \def\@tempa{& &} 
      \or \def\@tempa{&} 
      \or\def\@tempa{}\fi\@tempa 
{\rm(\theequation)}} 
 
\def\aeqno#1{\global\@fewtabfalse 
    \ifcase\@eqcnt \def\@tempa{& & &}\or \def\@tempa{& &} 
      \or \def\@tempa{&} 
      \or\def\@tempa{}\fi\@tempa 
{\rm(\theequation.#1)}} 
 
\def\label#1{ 
 \@bsphack\if@filesw {\let\thepage\relax 
   \def\protect{\noexpand\noexpand\noexpand}% 
\xdef\@gtempa{\write\@auxout{\string 
      \newlabel{#1}{{\@currentlabel}{\thepage}}}}}\@gtempa 
   \if@nobreak \ifvmode\nobreak\fi\fi\fi 
  \@esphack} 
 
\def\alabel#1#2{\label{#1}\global\@fewtabfalse 
    \ifcase\@eqcnt \def\@tempa{& & &}\or \def\@tempa{& &} 
      \or \def\@tempa{&} 
      \or\def\@tempa{}\fi\@tempa 
{\hbox to 2cm{\phantom{\rm(\theequation.#2)} 
\hfil} \hskip -2cm {\rm(\theequation.#2)}}} 

\def\clabel#1{\label{#1}\global\@fewtabfalse 
    \ifcase\@eqcnt \def\@tempa{& & &}\or \def\@tempa{& &} 
      \or \def\@tempa{&} 
      \or\def\@tempa{}\fi\@tempa 
{\hbox to 1cm{\phantom{\rm(\theequation)} 
\hfil}\hskip -1cm{\rm(\theequation)}}} 
\catcode`\@=12 
\begin{document} 
 
\begin{center} 
%%%%    \hfill hep-th/ 
%%%%    \vskip 5mm 
%%%%%%%%%%%%%%%%%%%%%%%%%%%%%%%%%%%%%%%%%%%%%%%%%%%%%%%%%%%% 
 
{\large\sc 
Universality and Ambiguities of the Conformal Anomaly} 
\vskip 6mm 
 
{\bf 
M. Asorey $^{a}$\!\! ,
%E-mail address: asorey@saturno.unizar.es}, 
%$\,\,\,\,\,\,\,\,\,\,\,\,\,\,\,\,\,\,\,\,\,\,$ 
 E. V. Gorbar $^{b,c}$ { and}\,\,  
%\footnote{E-mail address: gorbar@fisica.ufjf.br}, 
%$\,\,\,\,\,\,\,\,\,\,\,\,\,\,\,\,\,\,\,\,\,\,$ 
I.L. Shapiro $^{b}$ 
\footnote{On leave from Tomsk State Pedagogical University, 
Russia. %E-mail address: shapiro@fisica.ufjf.br 
} } 
 
\vskip 6mm 
 
{\small\sl 
(a) $\,\,$ Departamento de F\'{\i}sica Teorica, 
           Universidad de Zaragoza, 50009, Zaragoza, Spain 
 
(b) $\,\,$ Departamento de F\'{\i}sica -- ICE, 
           Universidade Federal de Juiz de Fora, MG, Brazil 
 
(c) $\,\,$ Bogolyubov Institute for Theoretical Physics, 
           Kiev, Ukraine} 
 
\vskip 8mm 
 
%%%%%%%%%%%%%%%%%%%%%%%%%%%%%%%%%%%%%%%%%%%%%%%%%%%%%%%%%%%% 
{\large\bf Abstract} 
\end{center} 
%%%%%%%%%%%%%%%%%%%%%%%%%%%%%%%%%%%%%%%%%%%%%%%%%%%%%%%%%%%% 

\begin{quotation} 
 
\noindent 
The one-loop structure of the trace anomaly is investigated 
using different regularizations and renormalization schemes: 
dimensional, proper time and Pauli-Villars. The universality 
of this anomaly is analyzed from a very  general perspective. 
The Euler and Weyl terms of the anomalous trace of the 
stress tensor are absolutely universal. The pure derivative 
$\,\square R$-term is shown to be universal only if the 
regularization  breaks conformal symmetry softly. If the 
breaking of conformal symmetry by the regularization method 
is hard the coefficient of this term might become arbitrary 
which points out the presence of an ambiguous 
$\,\int\sqrt{-g} R^2$-term in the effective quantum action. 
These ambiguities  arise in some prescriptions of dimensional 
and Pauli-Villars regularizations. We discuss the implications 
of these results for anomaly-induced inflationary scenarios 
and AdS/CFT correspondence. 
%%%%%%%%%%%%%%%%%%%%%%%%%%%%%%%%%%%%%%%%%%%%%%%%%%%%%%%%%%%%%%%% 
 
\end{quotation} 
\vskip 12mm 
 
%%%%%%%%%%%%%%%%%%%%%%%%%%%%%%%%%%%%%%%%%%%%%%%%%%%%%%%%%%%%%%%% 
%%%%%%%%%%%%%%%%%%%%%%%%%%%%%%%%%%%%%%%%%%%%%%%%%%%%%%%%%%%%%%%% 
 
\section{Introduction} 
 
\quad 
The anomalous violation of local conformal symmetry plays a 
central role in  many applications of  quantum field theory 
in curved space-times (see, e.g., \cite{duff94} for historic 
review). 
The trace anomaly is closely related to the renormalization 
group and encodes fundamental properties of the Quantum 
Field Theory in curved spaces which might have cosmological 
implications. In particular, the modified Starobinsky 
inflationary model is essentially based on the conformal 
anomaly (see \cite{asta} and references therein). The central 
idea of this model is the transition from a stable inflation 
in the UV regime to  an unstable one in the IR. This 
transition is due to the change of sign of the coefficient 
$\,a^\prime\,$ of the anomaly-induced 
$\,\int \sqrt{-g}R^2$-term in the effective action. 
 
The conformal anomaly  also played a fundamental role in the 
verification of 
the duality relation between string theory and gauge theories 
unveiled by Maldacena's conjecture \cite{scen}. 
In ${\cal N}=4$ SUSY gauge theories the $a'$ coefficient of 
the  $\sqrt{-g}R^2$ anomalous contribution vanishes if the 
contribution of  each field to  $\,a^\prime$ is the standard 
one. This means that the theory is incompatible with any 
inflationary scenario which is in agreement with the 
predictions coming from SUGRA in the large $\,N\,$ limit 
\cite{scen} and provides a further test of the AdS/CFT 
correspondence \cite{mald}. 
 
The conformal anomaly is a well known phenomena arising in 
the quantization of  conformally invariant classical field 
theories on  non-trivial gravitational backgrounds. In four 
dimensions conformal invariance requires that spinor, vector 
and scalar fields have to be massless. In scalar field theories 
conformal invariance also requires that the  $\xi$ parameter 
of the  $\, R\, \ph^2\,$ non-minimal interaction must have a 
fixed value $\xi=1/6$. 
 
The trace anomaly is generated by the renormalization of 
vacuum energy in the presence of an external space-time 
metric. In the functional integral approach it appears 
as a consequence of the lack of invariance of the functional 
measure under conformal transformations. This is reflected 
in perturbation theory by the appearance of new divergent 
terms in the effective action which depend on the background 
space-time metric and vanish for  Minkowskian backgrounds. 
They have to be removed 
by the corresponding counterterms in order to preserve the 
renormalizability of the theory (see, e.g. \cite{book} for 
an introduction). The renormalized effective action which 
emerges after this procedure is not conformally invariant. 
 
In  one-loop approximation the divergent terms are conformally 
invariant not only for the vacuum energy but also for the matter 
sector. Conformal invariance is only broken by the divergent 
coefficients which are of the form $\log M/\mu$ or $\mu^{n-4}$ 
depending on the regularization method.  However, at higher 
loops, even this marginal conformal invariance may be broken 
in the scalar field 
sector, because the $\be$-function for $\,\xi\,$ is not 
proportional to $\,\xi-1/6\,\,$ \cite{brwcol,hath}. Thus, 
beyond  one loop level the theory 
with  $\,\xi=1/6\,$ is non-renormalizable and  new divergent 
terms of the form  $\,\sqrt{-g}R^2\,$  appear in 
the vacuum sector. $\,\int\sqrt{-g}R^2\,$ counterterms have 
to be included to cancel higher loop divergences according to 
the prescription of any consistent renormalization scheme 
and the counterterms are not any more conformally invariant. 
 
In this paper, we shall only focus on  one-loop contributions 
to the conformal anomaly. The main question that we will 
address 
is what is the level of universality of the $\,a^\prime$ 
coefficient of 
the one-loop $\,\int\sqrt{-g}R^2$ anomalous term of the 
effective 
action. The conventional viewpoint establishes that this 
coefficient 
$\,a^\prime$  is scheme-dependent because 
$\sqrt{-g}R^2$ is a local term and its coefficient can 
always be 
fixed by a renormalization prescription without any 
restriction. 
However, most of the renormalization schemes considered 
in the 
literature give the same value for $\,a^\prime$, suggesting 
an unexpected universal behavior. The aim of the paper 
is to 
clarify this important issue. For this purpose, we shall 
consider 
different regularization schemes, looking for any possible 
explicit 
source of ambiguity for this coefficient $\,a^\prime$. 
The other 
terms of the one-loop anomaly are linked to  non-local 
terms 
of the effective action \cite{ddi} and are therefore 
universal. 
 
The paper is organized as follows. In section 2 a detailed 
discussion of the anomaly in dimensional regularization is 
presented. In particular, we show that, contrary to a widespread 
wisdom \cite{duff77,birdav}, the result of the calculation 
in dimensional regularization does not completely fix 
the value of the  $\,a^\prime\,$ coefficient. This 
ambiguity is equivalent to the freedom of adding 
a $\,\int\sqrt{-g}R^2$-term into the  action of the vacuum. 
In particular, it is possible to define a prescription based 
on  dimensional regularization, which provides a value for the 
$\,a^\prime\,$ coefficient which is identical to that 
generated by other regularization 
schemes like the point-splitting regularization \cite{christ} or 
the quantum effective action method \cite{bavi2}. In section 3 we 
apply the covariant cut-off regularization \footnote{Dimensional 
regularization has been used in Ref. \cite{bavi2} in the way 
suggested by Brown and Cassidy \cite{brocas}} of the proper-time 
integrals to derive the conformal anomaly in the framework 
of the effective action method \cite{bavi2}. In this 
way we get once more a result consistent with that 
of Ref. \cite{christ}. In section 4 we review the derivation 
of the effective action for the massive scalar field 
\cite{apco} and get identical results  that those of 
\cite{christ,bavi2} when taking the massless limit. 
 
In  section 5 we consider a covariant version of the Pauli-Villars 
regularization. In this regularization the result depends on 
whether we consider  $\,\xi=1/6\,$ or $\,\xi\neq 1/6\,$ for the 
massive Pauli-Villars regulator fields. In the first case the 
breaking of the conformal symmetry is soft and we get a value 
$\,a^\prime\,$ which is consistent with that of \cite{christ, 
bavi2}. But, if the regulators are not conformally coupled 
$\,\xi\neq 1/6\,$, we have a hard breaking of conformal symmetry 
and the coefficient $\,a^\prime\,$ becomes arbitrary and is highly 
dependent on the details of the regularization. This shows that 
the ambiguity in $\,a^\prime\,$ first unveiled in dimensional 
regularization also appears in a regularization scheme with fixed 
dimensionality $\,n=4$. Finally, in section 6 we analyze the 
consequences of this  ambiguity. Since the arbitrariness concerns 
only the initial point of the renormalization group flow there is 
still room for an inflationary scenario based on the conformal 
anomaly. However, the results might affect the conjectured AdS/CFT 
duality unless a similar ambiguity is found in the infrared 
renormalization of SUGRA in the AdS background.

%%%%%%%%%%%%%%%%%%%%%%%%%%%%%%%** * **%%%%%%%%%%%%%%%%%%%%%%%%%%%%%%%%%%% 
%%%%%%%%%%%%%%%%%%%%%%%%%%%%%%%** * **%%%%%%%%%%%%%%%%%%%%%%%%%%%%%%%%%%% 
%%%%%%%%%%%%%%%%%%%%%%%%%%%%%%%** * **%%%%%%%%%%%%%%%%%%%%%%%%%%%%%%%%%%% 
 
\section{Dimensional regularization} 
 
\quad 
To investigate the structure of one-loop vacuum quantum effects, 
it is sufficient to consider the simplest case: free quantum 
matter fields on a curved classical background $\,g_{\mu\nu}$. 
We shall consider five sorts of quantum fields. The actions of 
real scalar, Dirac fermion, and massless Abelian vector look 
as follows (here $d_i$ are conformal weights of the fields): 
\beq 
S_s = \int d^4 x\sqrt{-g}\,\left\{\,\frac12\,g^{\mu\nu} 
\partial_\mu\, 
\varphi\partial_\nu\varphi +\frac{\xi}{2}\,R\, 
\varphi^2\,\right\}\,,\,\,\,\,\,\,\,\,\,\, 
\,\,\,\,\, d_s = -1\,, 
\label{classical action} 
\eeq 
where $\,\xi=1/6\,$ in the four-dimensional, and 
$\,\xi=(n-2)/2(n-1)\,$ for the $\,n$-dimensional 
conformal case, 
\beq 
S_f=&\displaystyle
i\int d^4 x\sqrt{-g}\, {\bar \psi}\,\ga^{\mu}\na_\mu\,\psi 
\,,\qquad\qquad\  d_f= -\frac32\,;
\phantom{\Bigg ]} 
\label{mattero} 
\\ 
S_v=&\displaystyle
\int d^4 x\sqrt{-g}\,\left\{ - \frac14\,F_{\mu\nu}F^{\mu\nu} 
\right\}\,, 
\qquad\quad\  d_v=0\,. 
\label{matter} 
\eeq 
In addition, there are two examples of the higher derivative conformal 
fields: scalar \cite{rei} 
\beq 
S_{\rm hds}\,=\,\int d^4x\sqrt{-g}\,\phi\, 
\Big[\square^2 + 2R^{\mu\nu}\,\na_\mu\na_\nu 
 -\frac23\,R\square +\frac13\,(\na_\mu\,R)\,\na^\mu\,\Big]\phi\,, 
\qquad\quad  \,d_{\rm hds} = 0 
\label{4sca} 
\eeq 
and spinor \cite{cofe} 
\beq 
S_{\rm hdf}&=& 
i \int d^4 x\sqrt{-g}\, {\bar \chi}\,\gamma^{\mu} \left[ 
\na _{\mu}\square+R_{\mu\rho}\na^{\rho}-\frac{5}{12}\,R\na_{\mu} 
-\frac{1}{12}\,(\na_{\mu}R)\right]\,\chi 
\,,\qquad d_{hdf} = -\frac12\,. 
\label{tildeD} 
\eeq 
The functional integral over all five types of the fields leads 
to different coefficients in the conformal anomaly of the trace 
of the dynamical Energy-Momentum (stress) tensor, which is 
defined by 
\beq 
T^{\mu\nu}\,=\,-\frac{2}{\sqrt{-g}}\,\frac{\de S}{\de g_{\mu\nu}}\,. 
\label{tensor} 
\eeq 
Here $S$ is the classical action  of the matter fields 
(\ref{classical action}-\ref{tildeD}) with an additional 
metric dependent term $S_{\rm vac}=S_{\rm vac}[g_{\mu\nu}]$ which 
is necessary for the renormalizability of the theory 
\cite{birdav,book}. Since, quantum corrections of matter fields 
to the vacuum energy induce divergent terms which depend on the metric, it 
is necessary to consider a classical action for the metric field 
in order to absorb the UV divergences by renormalization of the 
corresponding couplings. In this way we not only renormalize the vacuum 
energy but also the stress tensor operator which is a composite 
operator. 
 
The vacuum action $\,S_{\rm vac}\,$ can be regarded as the action 
of an external metric field, which has not  been quantized. From 
this dynamical viewpoint it might include non-conformal or even 
non-local terms. However, since we are  interested only in 
quantization of matter fields one has to consider only the terms 
in $\,S_{\rm vac}\,$ which are absolutely necessary to keep the 
matter theory renormalizable. This means that in our case 
$\,S_{\rm vac}\,$ is made of only local terms because the 
non-local terms are not renormalized. In conformally invariant 
theories the  classical action associated to  one-loop 
counterterms $\,S_{\rm vac}\,$ can be chosen in such a way that it 
is conformally invariant up to surface terms \cite{book}, i.e. it 
is of the form $\,S_{\rm vac}\,$ (all notations are explained in 
the Appendix) \beq S_{\rm vac}\,=\,\int d^4x\sqrt{-g}\,\left\{ 
a_1C^2+a_2E+a_3{\square}R \,\right\}\,, \label{vacuum} \eeq where 
the last term is a pure surface term. With this choice of 
$\,S_{\rm vac}$, the classical conformal Noether identity is 
satisfied and hence the trace of (\ref{tensor}) is zero 
$T^\mu_\mu=0$. 
 
In dimensional regularization the counterterms have the form 
similar to (\ref{vacuum}) (see, e.g. \cite{birdav,book}) 
\beq 
\De S_{\rm vac} \,=\, -\,\frac{\mu^{n-4}}{n-4}\,\int d^nx\sqrt{-g} 
\,\left\{\be_1C^2+\be_2E+\be_3{\square}R \,\right\}\,. 
\label{vacuum infinite} 
\eeq 
Here the coefficients $\,\be_i\,$ are the 
$\be$-functions of the parameters $\,a_i\,$ 
of the classical action (\ref{vacuum}), and $\mu$ is the dimensional 
parameter of renormalization. 
 
The  trace anomaly has the form 
\beq 
T\,=\,<T^\mu_\mu> \,=\,-\,\frac{2}{\sqrt{-g}}\,g_{\mu\nu}\, 
\,\frac{\de \Ga_{\rm vac}}{\de g_{\mu\nu}}\,\neq 0\,. 
\label{old} 
\eeq 
Since the first calculation of the anomaly \cite{duff77}, $T$ 
has been computed in many different ways (see, e.g. \cite{birdav, 
duff94} for a list of references, and also 
\cite{scen} for the calculation based on the $\,AdS/CFT\,$ 
correspondence). All methods give an anomaly of the  form 
\beq 
T\,=\,aC^2+cE+a^\prime{\square}R 
\label{trace} 
\eeq 
with $\,a=\be_1\,$ and $\,c=\be_2\,\,$ 
\footnote{In the cosmological setting \cite{asta} 
different notations $\,w=\be_1$, $\,b=\be_2$ and $\,c=\be_3\,$ are 
being used.}. 
However, there is a disagreement in the last coefficient $a^\prime$. 
 
The conventional result of dimensional regularization \cite{duff77} 
 is $\,a^\prime=2\be_1/3\,$ while most of the other calculations 
agree that 
$\,a^\prime=\be_3$ (such that the anomaly is proportional to the 
trace of the coincidence limit of the $a_2(x,x^\prime)$ coefficient 
of the Schwinger-DeWitt expansion).  Notice that the dimensional 
regularization also predicts the 
value $\,a^\prime=\be_3$, but only in the case of global conformal 
(scale) transformations associated to the renormalization group 
\cite{tmf,book} (see also the discussion in \cite{duff94}, where 
this was mentioned as an {\it uneasy} point). In general,  when 
all five types of the fields are present with different degeneracies 
$\,\,N_s,\,N_f,\,N_v,\,N_{\rm hds},\,N_{\rm hdf}$, the expressions 
for the $\beta$-functions are the following 
\beq 
\begin{array}{lll} 
&\displaystyle\be_1 = \Big(\,\frac{1}{120}\,N_s + \frac{1}{20}\,N_f 
 + \frac{1}{10}\,N_v\,\Big) 
 \,-\, 
\Big(\,\frac{1}{15}\,N_{\rm hds} + \frac{1}{60}\,N_{\rm hdf}\,\Big)\,, 
\phantom{\Bigg (} 
\cr 
&\displaystyle \be_2  = -\,\Big(\,\frac{1}{360}\,N_s + \frac{11}{360}\,N_f 
 + \frac{31}{180}\,N_v\,\Big) 
 \,+ \,
\Big(\,\frac{7}{90}\,N_{\rm hds} + \frac{3}{40}\,N_{\rm hdf}\,\Big)\,, 
\phantom{\Bigg ]} 
\cr 
&\displaystyle\be_3 = 
\frac{1}{180}\,N_s + \frac{1}{30}\,N_f - \frac{1}{10}\,N_v 
 -\frac{2}{45}\,N_{\rm hds}-\frac{68}{45}\,N_{\rm hdf}\,. 
\end{array} 
\label{oef} 
\eeq 
It is easy to see that for the ordinary spinors $N_f$ and scalars 
$N_s$ the expressions $\,a^\prime=2\be_1/3\,$ and $\,a^\prime=\be_3$ 
do coincide, while for the three other fields the contribution provided 
by dimensional regularization differs from that arising in other 
regularization schemes. Because of this fact we shall consider the 
coincidence of  $N_f$ and $N_s$ contributions as an accidental fact. 
We shall try to better  understand  the origin of the difference between 
dimensional and other regularizations at  one-loop level. 
 
In ${\cal N}=4$ supersymmetric $SU(N)$ gauge theories 
$\,N_s=6(N^2-1),$ $\,\,N_f=2(N^2-1),$ $\,\,N_v=(N^2-1)$, 
$\,\,N_{\rm hds}\,=\,N_{\rm hdf}\,=\,0$, $\,$ also 
$\,\be_1=-\be_2\,$ and $\,a^\prime=0$. This is in 
agreement with the prediction of  the AdS/CFT correspondence 
as was verified by Henningson and Skenderis \cite{scen}. However, 
the cancelation of the coefficient  $\,a^\prime$ does not hold in 
dimensional regularization, because according to \cite{duff77} 
$\,a^\prime=2\be_1/3\,\neq 0$. It will be very interesting to 
investigate if there is an infrared regularization of SUGRA in 5-dimensions 
able to predict a different value for this coefficient.

This  value of  $\,a^\prime\,$ can be modified by 
adding a finite term 
\beq S_{+}\,=\,\al\,\int d^4x\sqrt{-g}\,R^2 
\label{R^2} 
\eeq 
to the classical action (\ref{vacuum}). In order to verify 
this, we perform a  conformal parametrization of the metric 
$$ 
g_{\mu\nu}={\bar g}_{\mu\nu}\,e^{2\sigma}\,, 
$$ 
where the fiducial metric ${\bar g}_{\mu\nu}$ has a 
fixed determinant. 
One can establish the following (very useful) identity, which 
is valid for any functional of the metric $A[g_{\mu\nu}]$: 
\beq 
 - \frac{2}{\sqrt{-g}}\,g_{\mu\nu} 
\frac{\delta A[g_{\mu\nu}]}{\delta g_{\mu\nu}} 
= - \frac{1}{\sqrt{-{\bar g}}}\,e^{- 4\sigma} 
\frac{\delta\,A[{\bar g}_{\mu\nu}\,e^{2\sigma}]}{\delta \sigma} 
\,\Big|_{{\bar g_{\mu\nu}}\rightarrow g_{\mu\nu}, 
\sigma\rightarrow 0}\,. 
\label{general identity} 
\eeq 
Together with the transformation rules derived in the Appendix, 
this gives 
\beq 
 - \frac{2}{\sqrt{-g}}\,g_{\mu\nu} 
\frac{\delta }{\delta g_{\mu\nu}}\,\int d^4x\sqrt{-g}\,R^2 
= 12\,{\square} R\,. 
\label{identity} 
\eeq 
Hence, after including the term (\ref{R^2}), the classical trace 
becomes non-zero $\,T^\mu_\mu=12\al \,{\square} R\,$ and the average 
value acquires exactly the same modification: instead of 
(\ref{trace}) we have 
\beq 
T\,=\, aC^2 + cE + (a^\prime + 12\al){\square} R\,. 
\label{trace1} 
\eeq 
 
The finite term (\ref{R^2}) of the classical action can be 
considered as a finite counterterm although strictly speaking, 
since there are not divergent terms of this type, one might 
think that it is not necessary to renormalize the theory. 
 
The anomaly can be derived by other regularizations methods. 
Two important examples are based on point-splitting regularization 
\cite{christ} and the non-local effective action 
\cite{bavi2,bavi3}. In both cases one is calculating 
the anomaly of the Noether identity for the local conformal 
transformation. Another relevant example is the 
$\ze$-regularization \cite{haw}. 
The conventional viewpoint  is that the prediction of 
dimensional regularization \cite{duff77,duff94} disagrees with 
other regularization methods which give a common result. 
As it was  pointed out in the Introduction, this 
ambiguity might be very relevant for cosmological 
applications. Hence it is worth to reconsider the method of \cite{duff77} 
and to perform a deeper analysis of the derivation of the anomaly by 
means of the dimensional regularization. 
In $n\neq 4$ dimensions the  renormalized one-loop 
action can be split into three parts 
\beq 
\Ga^{(1)}_{\rm ren} \,=\, 
S_{\rm vac}+{\bar \Ga}^{(1)} + \De S_{\rm vac}\,. 
\label{gamma} 
\eeq 
where $\,{\bar \Ga}^{(1)}\,$ is the non-renormalized one-loop quantum 
correction, which is divergent, conformal invariant and non-local, 
and  $\,\De S_{\rm vac}\,$ are the counterterms, which are  divergent 
(to guarantee the finiteness  $\Ga^{(1)}_{\rm ren}$)  and local, 
which remove all UV divergences by renormalizing the parameters 
$a_{1,2,3}$ of the classical action $S_{\rm vac}$ 
\footnote{Alternatively, 
one can define the vacuum action as 
$\int d^4x\sqrt{-g}({\tilde a}_1R_{\mu\nu\al\be}^2 + 
{\tilde a}_2R_{\mu\nu}^2 + {\tilde a}_3R^2)$ and renormalize the 
coefficients $\,{\tilde a}_1,\,{\tilde a}_2$ and ${\tilde a}_3$.}. 
The trace anomaly is generated by the fact that 
the counterterms $\,\De S_{\rm vac}\,$ are not conformal 
invariant. 
 
The source of arbitrariness come from the fact that the term $C^2$ 
can be extended for $n\neq 4$ in many different ways without 
violating  the minimal requirements of dimensional regularization. 
The explicit form of $C^2(d)$ depends on its dimension $d$ which 
in general can be of the form $\,\,d=n+\gamma (n-4)\,\,$, $\gamma$ 
being an arbitrary real parameter. The integration is 
$n$-dimensional and the arbitrariness introduced by $\gamma$ is 
purely algebraic. For any value of $\ga$ this counterterm is local 
and covariant, and cancels out the $C^2$-type divergence in 
(\ref{gamma}) The choice $\,\ga=-1\,$ has been used in 
\cite{duff77}, in order to preserve the algebraic form of the 
counterterm in $\int d^nx\sqrt{-g}C^2$, although there is no 
principle imposing such a requirement. With the choice 
$\,\ga=-1\,$, using Eq. (A17) one gets \beq T(C^2) 
\,=\,\frac{2}{\sqrt{-g}}\,g_{\mu\nu}\,\frac{\de}{\de g_{\mu\nu}} 
\,\frac{\mu^{n-4}}{n-4}\,\int d^nx\sqrt{-g}\,\be_1\,C^2(4) 
\Big|_{n\to 4} = C^2-\frac23\,{\square}R\,. \label{d=4} \eeq 
Another important observation is that, when the operator 
(\ref{general identity}) is applied to the superficial counterterm 
$\,\,\int d^nx\sqrt{-g}\,{\square} R\,,\,$ it gives, 
automatically, zero. In this way we arrive at the relation 
$a^\prime\equiv -2\be_1/3\,$ advocated in \cite{duff77}. 
 
On the other hand, the value $\,\ga=-1\,$ is just one of 
the many possible choices but any other value of $\,\ga\,$ 
is perfectly valid. Changing the 
value of $\ga$ we change the coefficient of the $\square R$-term 
in (\ref{d=4}). Therefore, from this 
point of view the dimensional regularization incorporates a certain 
degree of  arbitrariness in the determination of the coefficient $a'$. 
This change is equivalent to add the term 
$$ 
\frac{\gamma}{12}\,\int d^4x\sqrt{-g}\,R^2 
$$ 
to the classical action, with  $\ga=12\al$. Thus, the arbitrariness 
in the choice of $\,\De S_{\rm vac}\,$ 
is equivalent to the addition of a finite counterterm 
$\,\int d^4x \sqrt{-g}\,R^2$  into the classical action. 
 
One particular  value $\,\ga=0\,$ is somehow distinguished, 
because in this case the counterterm $\,\int d^nx\sqrt{-g}C^2(n)\,$ 
has, according to Eq. (A16), the simplest conformal 
transformation law. In fact, with  $\,\ga=0\,$ this 
term transforms exactly as in case of the global conformal 
transformation $\si={\rm const}$. In this case 
 the $\,{\square} R$-term of the anomaly is unrelated to 
the $\int \sqrt{-g}C^2$-type counterterm. It is worth to 
notice that in {\sl all} other regularization schemes there 
is no such dependence either. One can argue 
that this dependence is nothing but an artifact of the 
dimensional regularization. A remark  about the possibility of 
a similar arbitrariness in the dimensional regularization 
has been raised by Hawking \cite{haw}.

%%%%%%%%%%%%%%%%%%%%%%%%%%%%%%%%%%%%%%%%%%%%%%%%%%%%%%%%%%%%%%%% 
%%%%%%%%%%%%%%%%        %%%%%%%%%%%%%%%         %%%%%%%%%%%%%%%% 
%%%%%%%%%%%%%%%%%%%%%%%%%%%%%%%%%%%%%%%%%%%%%%%%%%%%%%%%%%%%%%%% 
\section{A Covariant Proper Time Cut-off Regularization} 
 
The relation between the anomaly and the heat kernel formalism 
for the effective action has been discussed by 
Vilkovisky et al in \cite{bavi2,bavi3}. The coefficient of 
the $\,\,\square R$-term in the anomaly agrees with the 
point-splitting regularization result \cite{christ} 
$\,a^\prime=\be_3$. Indeed, the calculations 
of \cite{bavi3} have been performed by means of  dimensional 
regularization within the formulation given by Brown and Cassidy 
\cite{brocas} which is different from the method used in 
\cite{duff77}. It is interesting to recalculate the anomaly 
by this method \cite{bavi3} but without using dimensional 
regularization. We shall use instead a covariant proper 
time cut-off regularization scheme to take care of the 
ultraviolet divergences. An advantage of this method is that 
it does not require an extension of the Weyl tensor to the 
dimensions $n\neq 4$, while allowing to preserve covariance. 
 
This regularization scheme employs the Schwinger proper time 
formalism 
\cite{proper} and an explicit cut-off on the lower limit of 
integration over proper time. Practical calculations are 
very close to the ones of Ref. \cite{bavi3}, hence we shall 
use the same notations. To calculate the $\,\square R$-term 
of the conformal anomaly, it is enough to know the effective 
action up to the second order in curvature. By using the heat 
kernel calculations of \cite{bavi2} for the operator 
$$ 
\hat{1}\square\,+\,\hat{P}\,-\,\frac{\hat{1}}{6}\,R\,, 
$$ 
(here $\,\hat{1}\,$ and $\,\hat{P}\,$ are the identity matrix 
and the $c$-number operator in the corresponding space) 
we obtain the following one-loop effective action in the 
covariant proper time cut-off regularization: 
\beq 
\begin{array}{lll} 
\displaystyle S\,\,=\,\,\frac12\, 
\int \limits_{1/\Lambda^2}^{\infty} 
\frac{ds}{s}\frac{\sqrt{-g}}{(4\pi s)^2} \,\tr &\Big\{\, 
\hat{1}\,+\,s\hat{P}\,+\,s^2\,\Big[ 
\,\hat{1}\,R_{\mu\nu}f_1(-s\square)R^{\mu\nu} 
\,+\,\hat{1}\,R\, f_2(-s\square)\,R\, 
\phantom{\Big ]} 
\\ 
&\displaystyle +\,\hat{P}f_3(-s\square)R + \hat{P}f_4(-s\square)\hat{P} 
+ {\cal R}_{\mu\nu}f_5(-s\square){\cal R}^{\mu\nu}\,\Big]\, 
\Big\}. 
\end{array} 
\label{action} 
\eeq 
$\La$ is a parameter of the covariant cut-off 
regularization (we shall call it $\,1/\Lambda^2$-regularization), 
the functions $\,f_i\,$ are defined in \cite{bavi2} and 
$\,{\cal R}_{\mu\nu}\,$ is defined as a commutator of two 
covariant derivatives acting on a field $\phi$ 
\beq 
[\nabla_{\mu}\nabla_{\nu} - \nabla_{\nu}\nabla_{\mu}]\,\phi 
\, =\, {\cal R}_{\mu\nu}\,\phi\,. 
\label{definition} 
\eeq 
 
In contrast to the conventional dimensional regularization the 
first two terms in (\ref{action}) give nonzero contributions to 
the cosmological constant and the Einstein term, which diverge as 
$\Lambda^4$ and $\Lambda^2$, respectively. However, by subtracting 
the corresponding counterterms, we can completely cancel these 
contributions. However, the situation is different for terms 
which are quadratic in curvature. By subtracting the appropriate 
local counterterms, one gets a finite effective action which 
contains non-local terms with $\,\,\ln (-\square/\mu^2)\,$, that 
generate the conformal anomaly. The actual calculation of the 
anomaly in the $\,1/\Lambda^2$-regularization is very simple 
because formulas (4.6)--(4.10) of \cite{bavi2} remain valid and 
formula (4.11), which defines the conformal variation of action, 
is replaced by \beq \delta_{\sigma}S\, =\, \frac{1}{2(4\pi)^2}\, 
\int d^4x \sqrt{-g}\, \tr \,\sigma \cdot 
\Big[\,\square\,\tilde{t}_1\Big(\frac{1}{\Lambda^2},\square\Big)\hat{P} 
+ 
\square\,\tilde{t}_2\Big(\frac{1}{\Lambda^2},\square\Big)R\hat{1} 
+ O(R^2)\,\Big]\,, \label{variation} \eeq where \beq 
\tilde{t}_1(s,\square)\, =\, \frac{f(-s\square)-1}{s\square}\,, 
\,\,\,\,\,\,\,\,\,\,\,\,\,\,\, \tilde{t}_2(s,\square)\, 
=\,\frac{f(-s\square)-1}{12\,s\square} - 
\,\frac{f(-s\square)\,-\,1\,-\,s{\square}/6}{2\,(s\square)^2}\,, 
\eeq \vspace{-3mm} \beq {\rm} 
\,\,\,\,\,\,\,\,\,\,\,\,\,\,\,\,\,\,\,\,\,\,\, f(-s\square)\, =\, 
\int_0^1 d\alpha \,\exp 
\left[\,\alpha(1-\alpha)\,s\square\,\right]\,. \eeq One can 
expand $\,\tilde{t}_1(1/\Lambda^2,\square)$ and 
$\tilde{t}_2(1/\Lambda^2,\square)\,$ into series in 
$\,1/\Lambda^2$. Retaining the leading term, we find \beq 
\tilde{t}_1(0,\square)\, =\, \frac{1}{6}\,,\,\,\,\,\,\,\,\,\,\,\, 
\,\,\,\,\,\,\,\,\,\,\, \tilde{t}_2(0,\square)\, =\, \frac{1}{180} 
\label{values} \eeq and, consequently, obtain the same result for 
anomaly that \cite{bavi2}, where the Brown--Cassidy regularization 
scheme \cite{brocas} was adopted. In this version of the 
dimensional regularization scheme there is no room for ambiguity. 
 
In order to illustrate how this scheme works in practice, 
let us perform 
the calculations for the particular case of a massless 
vector, where the conventional dimensional 
regularization and the method of \cite{brocas,bavi2} give 
different results. In this case the operator has the form 
\beq 
{\widehat H}\, =\, 
H^{\mu}_{\nu}\, =\, \delta^{\mu}_{\nu}\,\square 
+ {P}^{\mu}_{\nu} - \frac{1}{6}\,R\,\delta^\mu_\nu\,, 
\qquad {\rm where}\qquad 
P^\mu_\nu\,=\,-\,R^{\mu}_{\nu}+\frac{1}{6}\,R\,\de^\mu_\nu , 
\label{operator} 
\eeq 
and we also have to take into account the contribution to 
anomaly due to the ghosts 
\beq 
H_{gh}\, =\, \square\,. 
\label{ghosts} 
\eeq 
A  simple calculation shows that 
$\,\,\,a^\prime\,=\,\beta_3\,=\,-1/10\,,\,\,$ 
as it was expected from (\ref{oef}).

%%%%%%%%%%%%%%%%%%%%%%%%%%%%%%%%%%%%%%%%%%%%%%%%%%%%%%%%%%%%%%%% 
%%%%%%%%%%%%%%%% ****** %%%%%%%%%%%%%%% ******* %%%%%%%%%%%%%%%% 
%%%%%%%%%%%%%%%%%%%%%%%%%%%%%%%%%%%%%%%%%%%%%%%%%%%%%%%%%%%%%%%% 
\section{Anomaly and the massless limit of the massive 
field theory} 
 
\quad\quad 
Until now, we have considered only  massless fields. However, 
one can use an alternative approach and try to derive anomaly 
through a massless limit  of 
massive fields. The importance of this method is related to 
the fact that  quantum corrections coming from  massive 
fields are usually non-local and hence they can not be 
compensated by adding a local $\,\int\sqrt{-g}R^2$-term 
into the classical action. Furthermore, from the physical 
point of view, the concept of massless field is, in many 
cases, just an approximation valid in the high-energy region. 
 
The derivation of the effective action of massive 
fields has been performed recently in \cite{apco} using 
the asymptotic expansion for the trace of the heat kernel 
\cite{Avramidi,bavi2}. In \cite{apco} the calculations have 
been performed up to second order in curvature, which is 
sufficient for our purposes. The universal massless limit 
has been achieved for the three types of fields (scalars, 
fermions and vectors), and in all three cases the result 
is consistent with the point-splitting regularization 
$\,a^\prime=\be_3\,$ expression 
\cite{christ}. The result for the massive scalar field with a 
general non-minimal coupling $\,\,\xi$ up to second-order 
in the curvature part of the effective action is \cite{apco} 
\beq 
\begin{array}{c} 
\displaystyle 
{\bar \Ga}^{(1)}_{\rm scalar} 
\,=\,\frac{1}{2(4\pi)^2}\,\int d^4x \,g^{1/2}\, 
\left\{\,\frac{m^4}{2}\cdot\Big[\frac{2}{4-n} 
+\ln \Big(\frac{4\pi \mu^2}{m^2}\Big)+\frac32\Big]\, 
\right.\phantom{\Bigg ]} 
\\ 
\displaystyle\left. 
+\,\Big(\xi-\frac16\Big)\,m^2R\, 
\Big[\,\frac{2}{4-n} 
+\ln \Big(\frac{4\pi \mu^2}{m^2}\Big)+1\,\Big]\, 
\right.\phantom{\Bigg ]} 
\\ 
\displaystyle\left. 
+\,\frac12\,C_{\mu\nu\al\be} \,\Big[\,\frac{1}{30\,(4-n)} 
\,+\,\frac{1}{60}\ln \Big(\frac{4\pi \mu^2}{m^2}\Big)+k_W(a) 
\,\Big] C^{\mu\nu\al\be}\, 
\right. 
\phantom{\Bigg ]} 
\\ 
\displaystyle\left. 
+\,R \,\Big[\,\frac12\,\Big(\xi-\frac16\Big)^2\, 
\Big(\,\frac{2}{4-n} 
+ \ln \Big[\frac{4\pi \mu^2}{m^2}\Big]\,\Big) 
+ k_R(a)\,\Big]\,R\,\right\}\,. 
\end{array} 
\label{final scalar} 
\eeq 
The form factors are given by 
\beq 
k_W(a)\, = \,\frac{8A}{15\,a^4} 
\,+\,\frac{2}{45\,a^2}\,+\,\frac{1}{150}\,, 
\nonumber 
\eeq 
\vspace{-3mm} 
\beq 
\begin{array}{ll} 
k_R(a)\, = &\displaystyle 
\,A\Big(\xi-\frac16\Big)^2-\frac{A}{6}\,\Big(\xi-\frac16\Big) 
+\frac{2A}{3a^2}\,\Big(\xi-\frac16\Big) 
+\frac{A}{9a^4} 
\phantom{\Bigg ]} 
\\ 
&\displaystyle -\frac{A}{18a^2}+\frac{A}{144}+ 
+\frac{1}{108\,a^2} 
-\frac{7}{2160} + \frac{1}{18}\,\Big(\xi-\frac16\Big)\,, 
\end{array} 
\label{W} 
\eeq 
where 
\beq 
A \,=\,1-\frac{1}{a}\ln \frac{1+a/2}{1-a/2} 
\,\,\,\,\,\,\,\,\,\,\,\,\, {\rm and} \,\,\,\,\,\,\,\,\,\,\,\,\, 
a^2=\frac{4\na^2}{\na^2-4m^2}\,. 
\label{A} 
\eeq 
 
If the non-minimal coupling of the theory is conformal 
$\,\xi=1/6$, the massless limit $m\to 0$ is non-singular 
and we arrive at the expected $\,a^\prime=\be_3\,$ 
coefficient of the $R^2$-term 
\beq 
{\bar \Ga}^{(1)}(\xi=1/6,\,m\to 0) 
\,=\,-\,\frac{1}{12\cdot 180(4\pi)^2}\, 
\int d^4x \,g^{1/2}\,R^2 \,+ \dots \,. 
\label{anomaly-induced} 
\eeq 
Let us remark that in the limit $m\to \infty$ both form 
factors $\,k_W(a)\,$ and $\,k_R(a)\,$ tend to zero.

%%%%%%%%%%%%%%%%%%%%%%%%%%%%%%%%%%%%%%%%%%%%%%%%%%%%%%%%%% 
%%%%%%%%%%%%%%%%%%%%%%%%%%%%%%%%%%%%%%%%%%%%%%%%%%%%%%%%%% 
%%%%%%%%%%%%%%%%%%%%%%%%%%%%%%%%%%%%%%%%%%%%%%%%%%%%%%%%%% 
%%%%%%%%%%%%%%%%%%%%%%%%%%%%%%%%%%%%%%%%%%%%%%%%%%%%%%%%%% 
\section{Covariant Pauli-Villars regularization} 
 
\quad 
Pauli-Villars regularization is based on the introduction 
of a family of 
massive auxiliary fields which cancel the UV divergences of 
the theory in a 
covariant way and formally decouple from the original theory 
when their 
masses become  very large. The number of the regulator fields, 
their statistics and masses are chosen in such a way that all 
the divergences 
of the theory are canceled. In the singular limit when all 
their masses 
simultaneously tend to infinity, only covariant divergences 
appear and can 
safely be removed by appropriate renormalization schemes. In 
the Yang-Mills 
theories Pauli-Villars regularizations combined  with the 
introduction of 
higher derivatives regulators in the classical action provided 
non-perturbative continuum regularization  \cite{slavnov,asorey}, 
which is  very  useful to incorporate dynamical fermion fields 
while keeping 
all topological properties of gauge fields. In the present 
context we have only one-loop divergences, and therefore there 
is no need to apply higher derivatives regularization. 
For simplicity, we shall restrict our analysis to scalar fields, 
although the method also works for higher spin fields. 
 
The classical action of the single massless conformal 
field (\ref{classical action}) is replaced by the action 
\beq 
S_{\rm reg} = \sum_{i=0}^N\int d^4 x\sqrt{-g} 
\,\left\{\,\frac12\,g^{\mu\nu} 
\partial_\mu\, 
\varphi_i\partial_\nu\varphi_i +\frac{\xi_i}{2}\,R\, 
\varphi_i^2\,+ \frac{m^2_i}{2}\, 
\varphi_i^2\, \right\} 
\label{regaction} 
\eeq 
where the physical scalar field $\varphi$ (now labeled by 
$\varphi_0$) is conformally coupled ($\xi=\frac{1}{6}$), 
massless ($m_0=0$) and has bosonic statistics ($s_0=1$). 
The $N$ Pauli-Villars fields $\,$ $\varphi_i$ $\,$ 
($i=1,\dots,N$) $\,$ are massive $\, m_i=\mu_i M\neq 0$ 
and can have 
either bosonic $s_i=1$ or fermionic statistics $s_i=-2$. 
In the case of 
having Pauli-Villars fields with identical masses 
we shall incorporate the multiplicity as a factor in the 
coefficient 
$s_i$. This means that, in practice, for fields with fermionic 
statistics $\,s_i\,$ is a negative even integer whereas for 
those with bosonic statistics $\,s_i\,$ may be any natural 
number. We also assume, for the sake of completeness, 
that the Pauli-Villars regulators might 
have non-conformal couplings $\xi_i\neq \frac{1}{6}$. 
The regularized effective 
action of the massless scalar field with conformal 
coupling  $\xi=\frac{1}{6}$ is given by 
\beq 
{\bar \Ga}^{(1)}_{\rm reg}\,=\, 
\lim_{\Lambda\to \infty}\sum_{i=0}^N s_i {\bar \Ga}^{(1)}_{\rm i} 
\left(m_i,\xi_i,\Lambda\right)\,, 
\label{total} 
\eeq 
where $\Lambda$ is an auxiliar momentum cut-off. 
 
According to  general  prescriptions, all divergences in the 
ultraviolet cut-off $\Lambda$ are canceled out due to the Pauli-Villars 
conditions 
\beq 
&\displaystyle \sum_{i=1}^N \, {s_i}\,=\,-\,s_0\,=\,-\,1\,; 
\phantom{\Big ]}\,\,\, 
\alabel{pppp}{a} 
\cr 
&\displaystyle \sum_{i=1}^N \,s_i\, \mu_i^2\,=\,0\, ;\qquad%\qquad 
\sum_{i=1}^N \,s_i\,\left(\,\xi_i-\frac16\,\right)\,=\,0\,;
\phantom{\Big ]}
\alabel{pp}{b}\,\,\cr 
&\displaystyle \sum_{i=1}^N \,s_i\, \mu_i^4\,=\,0\, ;\qquad%\qquad 
\sum_{i=1}^N \,s_i\,\left(\xi_i-\frac{1}{6}\right)^2\,=\,0\,. 
\alabel{plog}{c}  
\cr
\nonumber 
\eeq 
%%%%%%%%%%%%%%%%%%%%%%%%%%%%%%%%%%%%%%%%%%%%%%%%%%%%%%%%%%%%%%% 
 
The first equation (\ref{pppp}.a) cancels out quartic divergences 
$\,\Lambda^4$, the second and third equations (\ref{pp}.b) cancel 
quadratic ones $\,\Lambda^2\,$ and the last two equations 
(\ref{plog}.c) are required to cancel logarithmic divergences 
$\,\,\log\,(\Lambda^2/m^2)$. 
A simple solution of these equations matching all these 
requirements is 
\beq 
&\displaystyle{s_1\,=\,1,\,\qquad s_2\,=\,4,\,\qquad 
s_3\,=-\,s_4 \,=\,\,s_5\,=-\,2}, 
\phantom{\Big ]} 
\cr 
&\displaystyle{\mu_1^2\,=\,4,\,\quad \mu_2^2\,=\,3, 
\,\quad \mu_3^2\,=\,1,\,\quad \mu_4^2\,=\,3,\,\quad \mu_5^2\,=\,4}, 
\phantom{\Big ]} 
\cr 
&\xi_1\,=\,4\,+\,\frac{1}{6},\,\quad \xi_2\, =\,3\,+ 
\,\frac{1}{6},\, \quad \xi_3\,=\,1\,+\,\frac{1}{6}, 
\,\quad \xi_4\,=\,3\,+\,\frac{1}{6},\,\quad \xi_5 
\,=\,4\,+\,\frac{1}{6}\,. 
\nonumber 
\eeq 
A compact expression of the effective action is obtained in the limit 
$\,M\to \infty\,$. 
In this limit the form factors $\,k_W(a)\,$ and $\,k_R(a)\,$ for 
the auxiliary fields vanish and the asymptotics of the remaining 
expression has the form 
\beq 
\begin{array}{ll} 
{\bar \Ga}^{(1)}_{\rm reg}(M) \,=&  \displaystyle \, 
\frac{1}{2\,(4\pi)^2}\,\int d^4x \,\sqrt{-g}\, 
\Big\{\,\frac{M^4\,\alpha}{2} 
\,+\, M^2\,R\,\beta\, + \,\Big(\delta-\frac{1}{1080}\Big)\,R^2\, 
\phantom{\Bigg ]} 
\cr 
&-\displaystyle \,\frac{1}{120}\,C^{\mu\nu\al\be}\,\ln\, 
\Big(\,\frac{\square}{{\rm e}^\gamma \,M^2}\,\Big)\, 
C_{\mu\nu\al\be}\,\Big\} 
\,\,+\,\,{\cal O} \Big(\, R^3 , \, \frac{1}{M}\,\Big)\,, 
\end{array} 
\label{totalfinal} 
\eeq 
where 
\beq 
&\displaystyle\alpha=\sum_{i=1}^N s_i \mu_i^4\ln\, \mu_i^2\,, 
\alabel{t}{a} 
\cr 
&\displaystyle\beta=\sum_{i=1}^N s_i\, \mu_i^2\, 
\Big(\,\xi_i-\frac16\,\Big)\ln\, \mu_i^2\,, 
\alabel{tt}{b} 
\cr 
&\displaystyle\gamma=\frac{46}{15}+\sum_{i=1}^N s_i\, \ln\, \mu_i^2\,, 
\alabel{ttt}{c} 
\cr 
&\displaystyle\delta=\sum_{i=1}^N s_i\, \Big(\,\xi_i-\frac16\,\Big)^2\, 
\ln\, \mu_i^2\,, 
\alabel{tttt}{d}\,\,\cr 
\nonumber 
\eeq 
The expression (\ref{totalfinal}) does not include pure divergences 
like $\square R$ or the Euler term. 
To obtain these terms would require a more involved 
calculation specifying the boundary conditions of the fields. 
The Euler term of the anomaly cannot be inferred from the above 
expression (\ref{totalfinal}), because it can be derived from 
higher order terms in the curvature  expansion \cite{bavi2} and 
hence is hidden in the $\,{\cal O}(R^3)\,$ term. 
Let us remind that here we are using the $\,{\cal O}(R^2)\,$ 
approximation described in the previous section. 
 
The Einstein  and cosmological constant terms of the regularized 
action (\ref{totalfinal}) are quartically and quadratically 
divergent in the large $\,M\,$ limit. Since they  are local, they 
can be renormalized in a way which is compatible with conformal 
invariance. This could be achieved by a suitable choice of the 
free parameters of the regularization (e.g. $\alpha=\beta=0$). 
The logarithmic divergent term cannot be canceled out by any 
choice of the parameters of the regularization and need also to be 
renormalized, but in this case any regularization prescription 
breaks conformal invariance. Upon renormalization, the regulating 
massive parameter $\,M\,$ is replaced by a renormalization scale 
$\mu$ which points out the existence of anomaly. The universality 
of the  coefficient multiplying the $\log \mu$ term (replacing 
$\log M$) in the effective action \pref{totalfinal}, can be 
understood in very simple terms. It is due to the soldering of 
this term  to that involving the non-local operator 
$\,\log(\Box/\mu^2)$, 
which cannot be renormalized and is, therefore, universal 
\cite{univamb,jll}. However, the $\,\int \sqrt{-g}R^2$ term has 
a finite coefficient $\,\delta-{1/ 1080}\,$ which is arbitrary 
and  not linked to any non-local operator. 
This term, unlike the $\,\int \sqrt{-g}C^2$ term whose coefficient 
$\gamma/120$ is also arbitrary,  provides a non-trivial 
contribution to the conformal anomaly unless we choose 
$\,\delta=1/1080$. 
 
Notice that $\delta$ vanishes if all Pauli-Villars field regulators have 
conformal couplings $\xi_i=\frac{1}{6},\,i=1,\dots N$. This 
condition is natural 
since in such a case the Pauli-Villars fields only break conformal 
invariance 
through their non-zero masses but do not have an extra breaking due to 
non-conformal couplings. In this way, we can understand why all other 
regularizations give a common value for the coefficient $a'=\frac{1}{180}$. 
In most regularizations the breaking of conformal symmetry is soft. 
However, if we introduce hard breaking regulators like scalar fields with 
non-conformal couplings we can have an arbitrary contribution $a'= 
\frac{1}{180}-6 \delta$ to the coefficient of $\square\,R$ term of the trace 
anomaly. These fields introduce logarithmic divergences in the 
$\,\int \sqrt{-g}R^2$ term which requires a non-trivial renormalization. 
The divergences are canceled out in our case by the Pauli-Villars condition 
(\ref{plog}), but they induce an arbitrary finite contribution which is 
encoded by $\delta$. A similar phenomenon occurs with any other soft 
regulator at two loops because in that case the coefficient of the 
$\,\int \sqrt{-g}R^2$ term becomes divergent as we send the masses of the 
regulators to infinity. Therefore, it requires a renormalization prescription 
which fixes its finite value at a given scale. One can notice that 
dimensional regularization obviously belongs to the same class, because 
the breaking of conformal symmetry is not soft in this case. 
 
There is no known principle excluding the use of hard 
regulators or prescribing 
the exclusive role of soft regulators. Therefore, 
what we have found is a real ambiguity in the one-loop 
contribution to the 
$\,\int \sqrt{-g}R^2$-term. This ambiguity is similar to that 
found in the renormalization of the Chern-Simons coefficient 
in 3-dimensional gauge 
theories \cite{univamb}. In fact, the above interpretation also 
holds in such a case. The Pauli-Villars regulators 
which induce the 
most general effective Chern-Simons coefficient 
introduce non-trivial logarithmic 
divergences (i.e. non-trivial beta functions of the 
gauge coupling constant) which cancel out under Pauli-Villars conditions. 
But those logarithmic divergent terms can leave a trace in the 
renormalized value of the  Chern-Simons coefficient. 
This provides for the 
first time a simple explanation in terms of  first principles 
of the striking phenomena observed in Chern-Simons theories 
\cite{univamb}. 
 
The existence of the ambiguity in the $\,\int \sqrt{-g}R^2$-term 
does not invalidate the field-theoretical foundations of the 
anomaly-induced inflation \cite{asta}. The beta function of the 
ambiguous $\,\int \sqrt{-g}R^2$-term in (\ref{totalfinal}) is zero 
at one loop level, therefore, the actual value of the 
corresponding coefficient can be fixed by imposing a 
renormalization condition\footnote{In fact, as it was already 
mentioned in the Introduction, a renormalization condition for 
this term is anyway necessary at higher loops because it might 
become logarithmically divergent.}. Although the quantum 
corrections to the finite $\sqrt{-g}R^2$-term are generally 
different for distinct regularizations, the renomalized theory may 
be the same, for the ambiguity can be compensated through the 
renormalization condition. Despite the one-loop $\,\beta$-function 
for the coefficient of the $\,\sqrt{-g}R^2$-term vanishes, the 
quantum effects induce a universal finite shift in $a^\prime$ from 
the UV to IR. In this sense the flow is universal and independent 
of the (scheme dependent) initial condition. The universal value 
of the finite shift for the conformal scalar is $\,\Delta 
a^\prime= -{1/1080}$ \footnote{This flow is very similar to the 
flow of the central charge in two-dimensional conformal theories 
where the Zamolodchikov $\,c$-theorem \cite{zamol} establishes a 
decreasing behaviour of the coefficient of the conformal anomaly 
from the  UV to the IR (see \cite{cardy,latorre} and references 
therein for  extensions to four dimensions).}. The existence of 
this shift is the basis of a mechanism of graceful exit of 
inflation in the modified Starobinsky scenario when supersymmetric 
particles of the MSSM model decouple. As a result of the change of 
the number of active degrees of freedom, the sign of the 
$a^\prime$ coefficient can change from positive in the UV (where 
the theory is supersymmetric) to the negative in the IR, when  
supersymmetry is broken. The physical requirements of the change 
of sign for the $\,a^\prime\,$ between the UV and IR regimes can 
be always achieved through the appropriate renormalization 
condition. One possible solution matching those requirements is to 
fix $a^\prime\approx\be_3$ for all fields \cite{asta}. 
 
If we assume that the ultimate reason behind the regulators 
is a more fundamental theory of space-time or string theory the 
type of regulator should be the same for all particles which 
survive at lower scales. Therefore, the nature of regulators 
(hard or soft) 
is fixed at very high energies and should affect all 
light particles 
in the same way. The fact that the choice $a^\prime=\be_3$ 
permits the existence of both inflation and graceful exit 
mechanisms in a most natural way suggests that the ultimate 
theory may be associated to soft breaking regulators rather 
than to hard breaking ones. 
 
In ${\cal N}=4$ SUSY gauge theories the existence of 
ambiguities in the coefficient $a^\prime$ implies that there 
should exist similar ambiguities in the infrared 
renormalization of supergravity 
in the bulk if the AdS/CFT correspondence is supposed to be 
robust under changes of renormalization schemes. Otherwise 
the correspondence will fail to match the  anomaly of the 
four-dimensional conformal field theory.

%%%%%%%%%%%%%%%%%%%%%%%%%%%%%%%** * **%%%%%%%%%%%%%%%%%%%%%%%%%%%%%%%%%%% 
%%%%%%%%%%%%%%%%%%%%%%%%%%%%%%%** * **%%%%%%%%%%%%%%%%%%%%%%%%%%%%%%%%%%% 
%%%%%%%%%%%%%%%%%%%%%%%%%%%%%%%** * **%%%%%%%%%%%%%%%%%%%%%%%%%%%%%%%%%%% 
 
\section{Conclusions} 
 
We investigated the universality problem of 
$\,\int \sqrt{-g}R^2$-terms 
induced  by the conformal anomaly in $4d$. The results 
may be summarized as follows: 
\vskip 1mm 
 
{\it i)\quad} We have verified  that the covariant proper-time 
cut-off renormalization scheme and the massless limit of 
massive fields lead  to   identical finite 
$\,\int \sqrt{-g}R^2$ contributions at  one-loop level. 
This result is perfectly consistent with that of point-splitting 
regularization method \cite{christ} and with the effective 
action approach of Ref. \cite{bavi2,bavi3}. This seems to suggest 
the existence of an unexpected universality of the 
$\square\, R$ term in the trace anomaly. 
\vskip 1mm 
 
{\it ii)\quad} Although it was claimed that the contribution 
to the coefficient $a'$ in dimensional regularization differs 
from what is obtained in  other renormalization schemes 
\cite{duff77} we have 
shown that there is no real disagreement between dimensional 
regularization and any other regularization scheme. Indeed, 
the usual way of applying the dimensional regularization 
\cite{duff77} can be easily and naturally generalized in such a 
way that the coefficient of this term and, correspondingly, the 
coefficient of the $\,\square R\,$-term in the trace anomaly 
become arbitrary. 
\vskip 1mm 
 
{\it iii)\quad} The ambiguous scheme-dependent 
$\,\int \sqrt{-g}R^2$-term can also be derived in a covariant 
Pauli-Villars regularization if the auxiliary massive scalar fields 
have an additional hard breaking of local conformal symmetry. 
This ambiguity is scale-independent and therefore can be 
fixed by the renormalization condition. However, the one-loop 
renormalization group flows for all three effective charges 
$\,a$, $\,c$, $\,a^\prime$ are universal. 
\vskip 1mm 
 
{\it iv)\quad} 
Despite the existing ambiguity in the $\,a^\prime$ coefficient, 
the renormalization group running of the  anomaly may be 
consistent with the requirements of the anomaly-induced 
inflation scheme. The ambiguity is scale-independent and therefore 
can be  fixed by the renormalization condition. 
 
Finally, it will be very interesting to analize if 
there are  infrared regularizations of SUGRA in 5-dimensions 
which generate, via the AdS/CFT correspondence, ambiguities 
similar to those we have found in the conformal anomaly 
in four dimensions. 
 
\bigskip 
\centerline{\large \bf Acknowledgments} 
We thank I. Avramidi for discussions. M.A. and Ed.G. 
thank the members of Departamento de F\'{\i}sica of UFJF 
for hospitality. I.Sh. is grateful for the warm 
hospitality of the Departamento de F\'{\i}sica Te\'orica 
of the University of Zaragoza, where this work was 
initiated. 
The work of M.A. has been partially supported by the Spanish 
MCyT grant FPA2000-1252. I.S. is indebted to CNPq for the 
permanent support and to FAPEMIG for the research grant. 
 
\vskip 12mm 
%%%%%%%%%%%%%%%%%%%%%%%%%%%%%%%%%%%%%%%%%%%%%%%%%%%%%%%%%%%%%%%% 
%%%%%%%%%%%%%%%%%%%%%%%%%%%%%%%%%%%%%%%%%%%%%%%%%%%%%%%%%%%%%%%% 
 
\noindent 
{\large\bf Appendix. Conformal transformations in $n$ 
dimensional space} 
\vskip 2mm 
 
In this appendix we summarize the conformal transformation 
laws of the most relevant geometric invariants in a general 
$\,n$-dimensional Riemannian manifold, which have 
been used in the paper. We use a condensed notations: 
$A_\mu^2=g^{\mu\nu}A_\mu A_\nu$, 
$(\nabla B)^2=g^{\mu\nu}\nabla_\mu B \nabla_\nu B$ 
and $\square B=g^{\mu\nu}\nabla_\mu \nabla_\nu B$. 
The  curvature tensors are defined by 
$$ 
{R^\al}_{\be\rho\tau} = \pa_\rho\Ga^\al_{\be\tau} - 
\pa_\tau\Ga^\al_{\be\rho} +  \Ga^\la_{\be\tau} \Ga^\al_{\la\rho} - 
\Ga^\la_{\be\rho} \Ga^\al_{\la\tau}\,, 
\eqno(A1) 
$$ 
$$ 
R_{\rho\si}={R^\al}_{\rho\al\si} \,, 
\,\,\,\,\,\,\,\,\,\,\,\,\,\,\,\,R=R^\al_\al\,. 
\eqno(A2) 
$$ 
Finally, 
closed parenthesis are used to bound the range of action of derivative 
operators, e.g. $\nabla_\mu B=(\nabla_\mu B)+B\nabla_\mu$. 
 
Consider the local conformal transformation 
$$ 
g^\prime_{\mu\nu}=g_{\mu\nu}\,e^{2\sigma}\,\,,\qquad 
\mbox{where} \qquad \,\sigma=\sigma(x)\,. 
\eqno(A3) 
$$ 
The transformation laws for the inverse metric and metric 
determinant have the form 
$$ 
g^{\prime\mu\nu}=g^{\mu\nu}\,e^{-2\sigma}\,,\, 
\,\,\,\,\,\,\,\,\,\,\,\,\,\,\,\,g^\prime=g\,e^{2n\sigma} 
\,,\,\,\,\,\,\,\,\,\,\,\,\,\,\,\,\,\,g=\det(g_{\mu\nu})\,. 
\eqno(A4) 
$$ 
For the Christoffel symbols we have 
$$ 
{\Ga^\prime}^\la_{\al\be} = \Ga^\la_{\al\be} + 
\de^\la_\al (\na_\be\si)+\de^\la_\be (\na_\al\si) 
-g_{\al\be} (\na^\la\si) 
\eqno(A5) 
$$ 
and for the curvatures 
\def\theequation{{A6}} 
\beq 
\begin{array}{lll} 
\displaystyle {R^{\prime\al}}_{\be\mu\nu}=&\displaystyle{R^\al}_{\be\mu\nu} 
+\de^\al_\nu(\na_\mu\na_\be\si)-\de^\al_\mu(\na_\nu\na_\be\si) 
+g_{\mu\be}(\na_\nu\na^\al\si)-g_{\nu\be}(\na_\mu\na^\al\si) 
\phantom{\Big ]} 
\cr 
&\displaystyle 
+\de^\al_\nu g_{\mu\be}(\nabla \si)^2 
-\de^\al_\mu g_{\nu\be}(\nabla \si)^2 
+ \de^\al_\mu(\na_\nu\si)(\na_\be\si) 
-\de^\al_\nu(\na_\mu\si)(\na_\be\si)  
\phantom{\Big ]}
\cr
&\displaystyle
+g_{\nu\be}(\na_\mu\si)(\na^\al\si) 
-g_{\mu\be}(\na_\nu\si)(\na^\al\si)\, ,
\phantom{\Big ]} 
\cr 
\end{array} 
\label{A6} 
\eeq 
\vskip 1mm 
$$ 
R^\prime_{\mu\nu}=R_{\mu\nu}-(n-2)(\na_\mu\na_\nu\si) 
-g_{\mu\nu}(\square \si)+(n-2)(\na_\mu\si)(\na_\nu\si) 
-(n-2)g_{\mu\nu}(\na\si)^2\,, 
\eqno(A7) 
$$ 
\vskip 1mm 
$$ 
R^\prime=e^{-2\sigma}\,\Big[\,R-2(n-1)(\square \sigma) 
-(n-1)(n-2)(\nabla \sigma)^2\,\Big]\,. 
\eqno(A8) 
$$ 
 
The Weyl tensor in $n$ dimensions is defined as 
$$ 
\begin{array}{ll} 
\displaystyle{ 
C_{\al\be\mu\nu}}=&\displaystyle{R_{\al\be\mu\nu} 
+\frac{1}{n-2}\,(g_{\be\mu}R_{\al\nu}-g_{\al\mu}R_{\be\nu} 
+g_{\al\nu}R_{\be\mu}-g_{\be\nu}R_{\al\mu})} 
\phantom{\Big ]}\cr 
&\displaystyle{ +\frac{1}{(n-1)(n-2)}\,R\,(g_{\al\mu}g_{\be\nu}- 
g_{\al\nu}g_{\be\mu})\,.} 
\end{array} 
\eqno(A9) 
$$ 
It is  traceless $C^\al_{\,\,\,\mu\al\nu}=0$ and transforms trivially, 
$$ 
C^{\prime\al}_{\,\,\,\,\be\mu\nu}\,=\,C^\al_{\,\,\,\be\mu\nu}\,, 
\qquad {\rm i.e.}\qquad 
C^\prime_{\al\be\mu\nu}\,=\,e^{2\si}\,C_{\al\be\mu\nu}. 
\eqno(A10) 
$$ 
 
The transformation of the curvature-square scalars are 
\def\theequation{{A11}} 
\beq 
\begin{array}{lll} 
\displaystyle\sqrt{-g^\prime}\,R^{\prime\,2}_{\mu\nu\al\be} 
=&\displaystyle\sqrt{-g}\,e^{(n-4)\sigma}\,\Big\{\,R^2_{\mu\nu\al\be} 
+8R^{\mu\nu}(\na_\mu\si)(\na_\nu\si) 
-8R^{\mu\nu}(\na_\mu\na_\nu\si) 
\phantom{\Bigg ]} 
\cr 
&\displaystyle -4R(\na\si)^2 + 4(\square \si)^2 
+2(n-2)\Big[2(\na_\mu\na_\nu\si)^2 
+4(\square \si)(\na\si)^2
\cr
&\displaystyle
-4(\na_\mu\na_\nu\si)(\na^\mu\si)(\na^\nu\si)
+(n-1)(\na\si)^4\Big]\,\Big\}
\phantom{\Bigg ]} \,, 
\cr 
\end{array} 
\label{A11} 
\eeq 
\vskip 1mm 
\def\theequation{{A12}} 
\beq 
\begin{array}{lll} 
\displaystyle\sqrt{-g^\prime}\,R^{\prime\,2}_{\mu\nu} 
=&\displaystyle \sqrt{-g}\,e^{(n-4)\sigma}\,\Big\{\,R^2_{\mu\nu} 
- 2R(\square \si) 
+(3n-4)(\square \si)^2 
- (n-2)[2R^{\mu\nu}(\na_\mu\na_\nu\si)
\phantom{\Big ]} 
\cr 
&\displaystyle -2R^{\mu\nu}(\na_\mu\si)(\na_\nu\si)
-(n-2)(\na_\mu\na_\nu\si)^2-(n-1)(n-2)(\na\si)^4 
\phantom{\Bigg ]}\cr 
&\displaystyle +2R(\na\si)^2+2(n-2)(\na_\mu\na_\nu\si)(\na^\mu\si)(\na^\nu\si) 
-2(2n-3)(\square \si)(\na\si)^2] 
\,\Big\}\,.\cr 
\end{array} 
\label{A12} 
\eeq 
\vskip 1mm 
$$ 
\sqrt{-g^\prime}\,R^{\prime\,2} 
=\sqrt{-g}\,e^{(n-4)\sigma}\,\Big[\,R-2(n-1)(\square \sigma) 
-(n-1)(n-2)(\nabla \sigma)^2\Big]^2\,. 
\eqno(A13) 
$$ 
The square of the Weyl tensor 
$$ 
C^2(n)=C_{\al\be\mu\nu}C^{\al\be\mu\nu}=R^2_{\mu\nu\al\be} 
-\frac{4}{n-2}\,R^2_{\mu\nu}+\frac{2}{(n-1)(n-2)}\,R^2 
\eqno(A14) 
$$ 
transforms as 
$$ 
\sqrt{-g^\prime}\,C^{\prime\,2}_{\mu\nu\al\be} 
=\sqrt{-g}\,e^{(n-4)\sigma}\,C^2_{\mu\nu\al\be} \,. 
\eqno(A15) 
$$ 
One can establish a simple relation between 
$$C^2(4)=R_{\mu\nu\al\be}^2 - 2R_{\al\be}^2 + 
1/3\,R^2$$ 
and $C^2(n)$: 
$$ 
C^2(4)=C^2(n)+\frac{2(n-4)}{n-2}\,R^2_{\mu\nu} 
-\frac{(n-4)(n+1)}{3(n-1)(n-2)}\,R^2\,. 
\eqno(A16) 
$$ 
A similar relation can be derived between $C^2(n)$ and 
$C^2(n+\ga[n-4])$, where $\ga$ is an arbitrary parameter. 
Another important combination is 
$$ 
E = R_{\mu\nu\al\be}R^{\mu\nu\al\be} 
- 4 \,R_{\al\be}R^{\al\be} + R^2\,. 
\eqno(A17) 
$$ 
 $E$ is the topological density of the Gauss-Bonnet topological term 
$\,\,\int d^n \sqrt{-g}\,E\,\,$ in  $n=4$. But  this term 
does not contribute to the propagator of gravitons (traceless and 
completely transverse modes) in any dimension. The conformal 
transformation of the Gauss-Bonnet term reads 
\def\theequation{{A18}} 
\beq 
\begin{array}{llll} 
\sqrt{-g^\prime}\,E^\prime 
\,=&\,\sqrt{-g}\,e^{(n-4)\sigma}\,E\,\,+\,\, 
\sqrt{-g}\,e^{(n-4)\sigma}\,(n-3)\, 
\Big[\,8R^{\mu\nu}(\na_\mu\na_\nu\si) 
\phantom{\Bigg ]} 
\cr 
& -8R^{\mu\nu}(\na_\mu\si)(\na_\nu\si) 
- 2(n-4)R(\na\si)^2 
- 4(n-2)(\na_\mu\na_\nu\si)^2 
\phantom{\Big ]} 
\cr 
&+ 4(n-2)(\square \si)^2 
+ 8(n-2)(\na_\mu\si)(\na_\nu\si)(\na^\mu\na^\nu\si) 
-4R(\square \si) 
\phantom{\Bigg ]} 
\cr 
&+4(n-2)(n-3)(\square \si)(\na\si)^2 
+(n-1)(n-2)(n-4)(\na\si)^4\,\Big] \,. 
\cr 
\end{array} 
\label{app18} 
\eeq 
The last remaining invariant is the surface term with the 
following transformation rule: 
\def\theequation{{A19}} 
\beq 
\begin{array}{llll} 
&\displaystyle \sqrt{-g^\prime}\,({\square}^\prime R^\prime) 
=\sqrt{-g}\,e^{(n-4)\sigma}\,\Big[\,({\square} R) 
-2(n-1)(\square^2 \si)-(n-1)(n-2)\square (\nabla \si)^2 
\phantom{\Bigg ]} 
\cr 
&\displaystyle 
+(n-6)(\na^\mu\si)(\na_\mu R)-2R(\square \si) 
-2(n-1)(n-6)(\na^\mu\si)(\na_\mu\square\si) 
\phantom{\Big ]} 
\cr 
&\displaystyle 
-(n-1)(n-2)(n-6)(\na^\mu\si)\na_\mu (\na\si)^2 
+4(n-1)(\square \si)^2 -2(n-4)R(\na\si)^2 
 \phantom{\Bigg ]} 
\cr 
&\displaystyle 
+2(n-1)(3n-10)(\square \si)(\na\si)^2 
+2(n-1)(n-2)(n-4)(\na\si)^4\,\Big] \,. 
\end{array} 
\label{a19} 
\eeq

%%%%%%%%%%%%%%%%%%%%%%%%%%%%%%%%%%%%%%%%%%%%%%%%%%%%%%%%%%%%%%% 
%%%%%%%%%%%%%%%%%%%%%%%%%%%%%%%%%%%%%%%%%%%%%%%%%%%%%%%%%%%%%%% 
 
\newpage 
%\vskip 20mm 
 
%%%%%%%%%%%%%%%%%%%%%%%%%%%%%%%%%%%%%%%%%%%%%%%%%%%%%%%%%%%%%%%%%%%%%%% 
  %%{99} 
 
\end{document}